\documentclass[preprint,12pt]{elsarticle}

\usepackage{graphicx}
\usepackage{natbib}
\usepackage{amssymb}
\usepackage{epsfig}
\usepackage{graphics}

\journal{International Journal of Heat and Mass Transfer}

\begin{document}

\begin{frontmatter}

\title{Non-Boussinesq Rolls in 2d Thermal Convection}

\author[label1,label2]{C. M\'alaga}
\author[label1]{F. Mandujano}
\author[label1]{R. Peralta-Fab}
\author[label1]{C. Arzate}

\address[label1]{Departamento de F\'isica, Facultad de Ciencias,
Universidad Nacional Aut\'onoma de M\'exico,04510,  M\'exico D.F}
\address[label2]{Email address for correspondence: cmi@fciencias.unam.mx}

\begin{abstract}
A study of convection in a circular two dimensional cell is presented. The system is heated and cooled at two diametrically opposed
points on the edge of the circle, which are parallel or anti-parallel to gravity. The latter's role in the plane of the cell can be changed by tilting the cell.
When the system is in a horizontal position, a non-trivial analytic solution for the temperature distribution of the quiescent fluid
can be found. For a slight inclination, the projection of gravity in the plane of the cell is used as a perturbation parameter in the
full hydrodynamic description, as the Boussinesq approximation is inadequate. To first order, the equations are solved for the
stationary case and four symmetrical rolls become apparent, showing that a purely conductive state is impossible if gravity -however small- is 
present; an approximate closed analytical expression is obtained, which  describes the four convection rolls. Further analysis is
done by a direct numerical integration. Comparison with preliminary observations is 
mentioned.
\end{abstract}


\end{frontmatter}

\section{Introduction}
\begin{figure}
\centering
\includegraphics[scale=0.4]{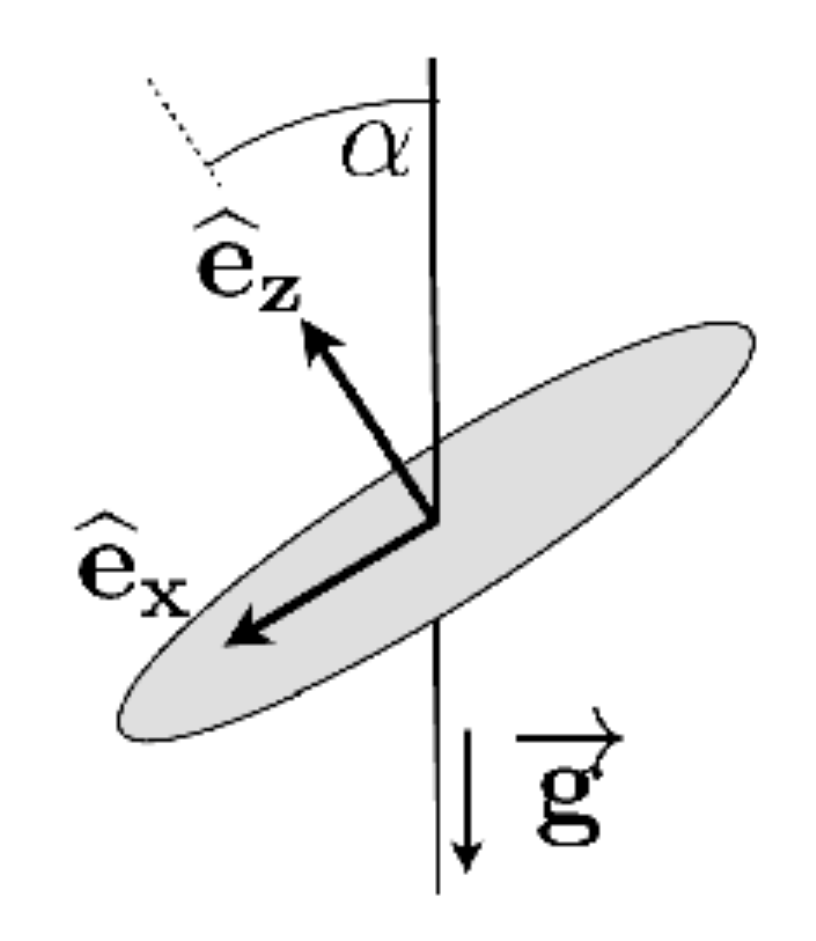}
\caption{}\label{D}
\end{figure}

\begin{figure}
\centering
\includegraphics[scale=0.3]{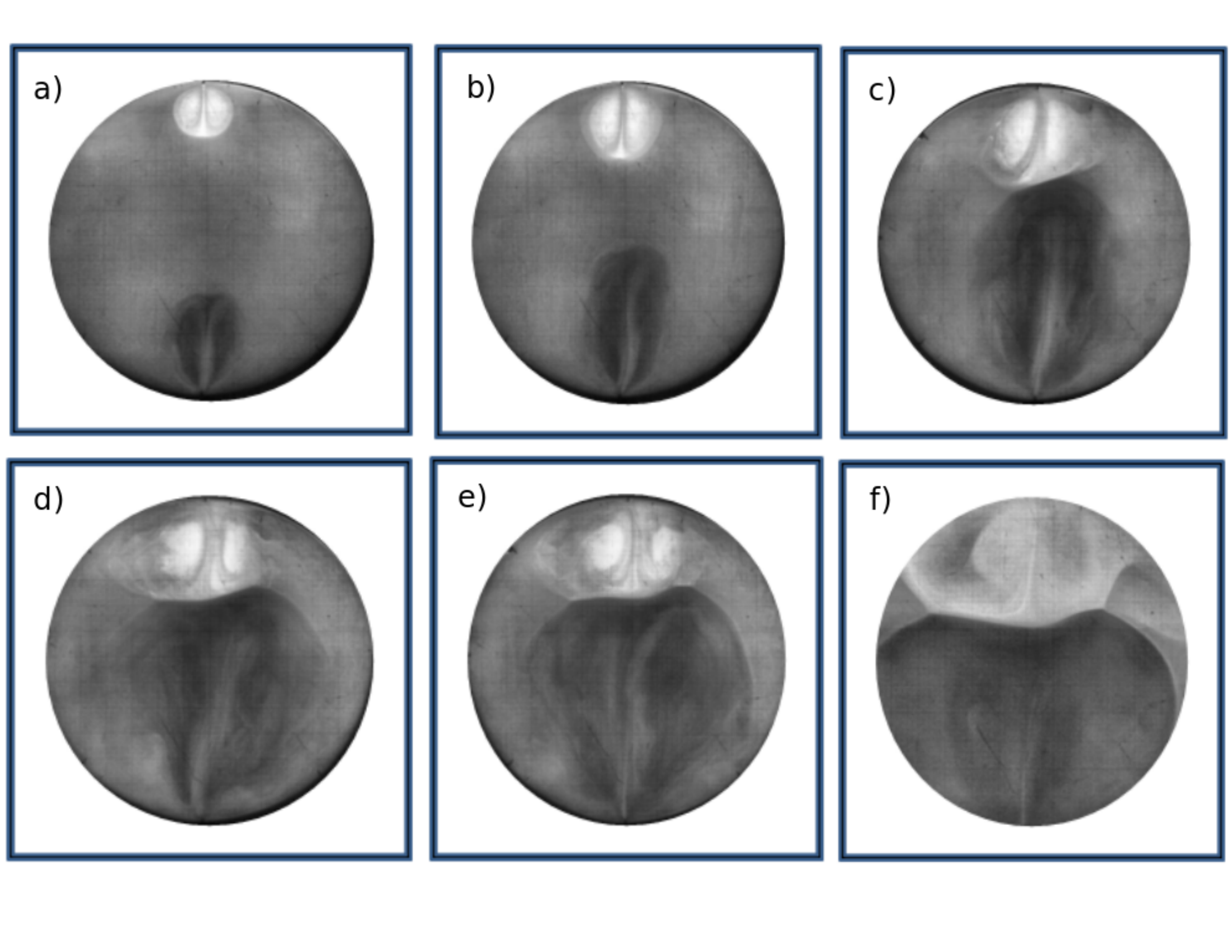}
\caption{A sequence of snapshots covering the first 15 minutes of the experiment, in the transient state, showing the development of the convection rolls}\label{E}
\end{figure}

Convection is usually studied in systems consisting of two parallel and horizontal plates at different fixed temperatures, gravity being
 perpendicular to the plates \citep{boden,lappa}. Such systems exhibit convective flows when the bottom plate is at a higher temperature
 than the top plate and the temperature difference exceeds some critical value; such flows might not be unique \citep{LT}. Below that
 value a purely conductive state is observed; this state can only be obtained if the plates are horizontal and at uniform temperatures. It
 appears that in any other geometry or temperature distribution the hydrostatic conditions cannot be satisfied, resulting in a convective
 flow \citep{landau}, even if the system is cooled at the bottom and heated at the top.\\
Analytically, convection is often studied using the Boussinesq approximation \citep{Chandra,gershuni}, in which the density variations are
 only taken into account through the body force (gravity) in the momentum equation. In fact, this common approach assumes that the density
 is only related to the temperature, and not to the pressure, or implicitly, the latter is considered constant. For a simple homogeneous
 fluid, any set of equations of state, such as those for an ideal gas or its generalizations (van der Waals, hard spheres, etc.), connect
 all three variables, unless justified isobaric suppositions are introduced explicitly.\\
In the geometry considered here, the body force can be made arbitrarily small, so that the Boussinesq approximation can no longer be used,
 forcing a full hydrodynamic description, as well as a fine tuning of the parameters involved in the momentum, mass, and energy
 conservation laws. 
We consider a cell in which a Newtonian fluid is restricted to move in a plane, within a circle. At two diametrically opposed points, on
 the rim, a heat source and a heat sink are located at fixed temperatures. By tilting the cell with respect to the vertical (defined by
 gravity), one can change the body force (see figure \ref{D}). A horizontal cell ($\alpha=0$) corresponds to a fluid subjected only to a temperature
 gradient and no external forces.
An experimental observation of the formation of a four-roll flow structure on a vertical quasi-two dimensional convection cell is shown in figure \ref{E}. 
This experiment corresponds to a cylindrical circular cell with a radius of $94mm$ and a thickness (height) of $1.9 mm$ in the vertical
position.
 The working fluid is a mixture of water and Kalliroscope\texttrademark (less than $5$\%), to help visualization of the flow structure. The system is heated at the bottom ($48^{\circ} C$) and cooled at the top ($8^{\circ} C$) through copper wires, $0.8mm$ in
 diameter, inserted so that their tips protrude $1.0mm$ into the cell, in full contact with the working substance. The picture represents
 a sequence of snapshots of the first 15 minutes of the experiment. The four-roll pattern varies slightly for hours.\\
Here, the parameters characterizing the fluid are the shear and bulk viscosity coefficients $\eta$ and $\mu$, and the thermal conductivity
 coefficient $\kappa$;  thermodynamics introduces the universal gas constant $R$, and the specific heat at constant volume $c_v$.  Additionally,
 the radius of the cell is $a$, its width is $h$ ($h \ll a$), and the mass it contains is $m$; an effective gravity, defined below, depends on the
 cell's tilt. This quantity appears naturally in place of the body force; if it is small, a perturbation expansion can be proposed for the
 velocity, density, temperature, and pressure fields in the 2d Navier-Stokes equations, the mass and energy conservation laws, and the two
 thermodynamic equations of state. At zeroth order, a static state is found, while higher orders exhibit the destabilizing effect of pressure
 variations.

\section{The Model}
In what follows, a homogeneous simple fluid, whose properties are assumed to be known, is restricted to two dimensions. The two equations of
 state are assumed to be known, one relating mass density $\rho$, pressure $p$ and temperature $T$, the other relates the internal energy
 density $e$ to the extensive thermodynamic variables. An ideal-like fluid satisfies
\begin{equation}
 p  = R \rho T,
\label{edo}
\end{equation}
\noindent where $R$ is a constant; for a perfect gas it would be the universal gas constant per unit mass. Accordingly, the internal energy is
 considered to be a function of temperature only. The continuity and momentum equations \citep{landau,batchelor} are given by
\begin{equation}
 D_t \rho + \rho \nabla \cdot {\bf u} = 0, 
\end{equation}
\noindent and
\begin{equation}
 \rho D_t {\bf u} = - \nabla p + \eta  \nabla^2 {\bf u} + \mu \nabla (\nabla \cdot {\bf u}) + \rho {\bf g}_{e},
\end{equation}
\noindent where ${\bf u}$ is the two-dimensional velocity field, the operator $D_t \equiv \partial_t +  {\bf u} \cdot \nabla$, and 
$\rho{\bf g}_{e}$ is the body force in the plane of the cell. If the cell is vertical ($\alpha=\pi/2$), ${\bf g}_{e}$ is simply the acceleration
 of gravity, $g$, times a unit vector pointing downwards; here, this is chosen as the $x$-direction. If the cell is horizontal, gravity plays no
 role and ${\bf g}_{e}={\bf 0}$. In general ${\bf g}_{e}=\hat{\bf e}_{x} g \sin \alpha$, $\alpha$ being the angle between the unit vector
  normal to the plane of the cell $\hat{\bf e}_{z}$ and the direction of gravity {\bf g}, (see figure \ref{D}).\\
The energy equation \citep{landau} is
\begin{equation}
c_v \rho D_t T  = - p \nabla \cdot {\bf u} + \kappa  \nabla^2 T + \frac{1}{2} \eta \mathring{(\nabla{\bf u}) }^2 + 
\mu (\nabla \cdot {\bf u} )^2,
\end{equation}
\noindent where $\mathring{(\nabla{\bf u}) }$ stands for the symmetrical traceless tensor derived from $\nabla{\bf u}$. 
Boundary conditions are the usual no-slip and zero heat flux on the circle (the edge of the cell), except for two points on the circle where the
 temperature is set to constant values, $T_h$ for hot, and $T_c$ for cold ($T_c<T_h$). Namely, in plane polar coordinates
 \begin {eqnarray}
{\bf u}={\bf 0},\ \  \partial_r T = 0, \qquad\mbox{at } r=a
\label{bc}
\end{eqnarray}
\noindent where the temperature's partial derivative is in the radial direction. At two diametrically opposed points, at say $\theta=\theta_0$,
 and $ \theta=\theta_0+\pi$, depending on which source is set on the upper point ($\theta_0=0$ for the usual hot below arrangement)
\begin {eqnarray}
T=T_c,\ \ T=T_h.
\end{eqnarray}
The dimensionless equations are obtained by rescaling variables with $g$, $m$, $a$, and $T_m = \frac{1}{2}(T_c+T_h)$. Considering dimensional
 variables with an asterisk, the rescaling relations are:
\begin{eqnarray}
 {\bf x}^*  = a {\bf x},& \ \ &t^* = t \sqrt{a/g}, \ \  T^* = T  T_m ,
 \nonumber  \\
  \rho^* =  \rho m/a^3,&  \ \ &{\bf u}^* = {\bf u}\sqrt{ag}.\nonumber
\end {eqnarray}
The resulting dimensionless governing equations are:
 \begin{eqnarray}
D_t \rho&=&- \rho \nabla \cdot {\bf u}, \label{cont} \\
\rho D_t {\bf u}&=&- \frac{1}{\mathcal{ M}^2} \nabla  (\rho T) + \mathcal{G} \left[ \nabla^2 {\bf u} + \frac{\mu}{\eta}  \nabla (\nabla \cdot {\bf u})
 \right] + \rho \sin(\alpha) \hat{\bf e}_x, \\
\rho D_t T& =&- \mathcal{ R } \rho T \nabla \cdot {\bf u} +  \frac{\mathcal{G}}{\mathcal{P}} \nabla^2 T +  \mathcal{ S } \left[ \frac{1}{2}
 \mathring{(\nabla{\bf u}) }^2 + \frac{\mu}{\eta} (\nabla \cdot {\bf u} )^2 \right], \label{ener} 
\end{eqnarray}
\noindent where the pressure has been eliminated using the equation of state, eqn.(\ref{edo}). The dimensionless parameters are a Prandtl
 number $\mathcal{P}=c_v \eta/\kappa$, $\mathcal{G} = a^{3/2} \eta / (m g^{1/2})$, the viscosity ratio $\lambda=\mu/\eta$, $\mathcal{ M}^{-2}= 
R T_m/(ag)$, $\mathcal{ R } = R / c_v$ and $\mathcal{S}= \eta a^{3/2} g^{1/2}/(c_v m T_m)$.\\
\section{Static Solution}
\begin{figure}
\centering
\includegraphics[scale=0.19]{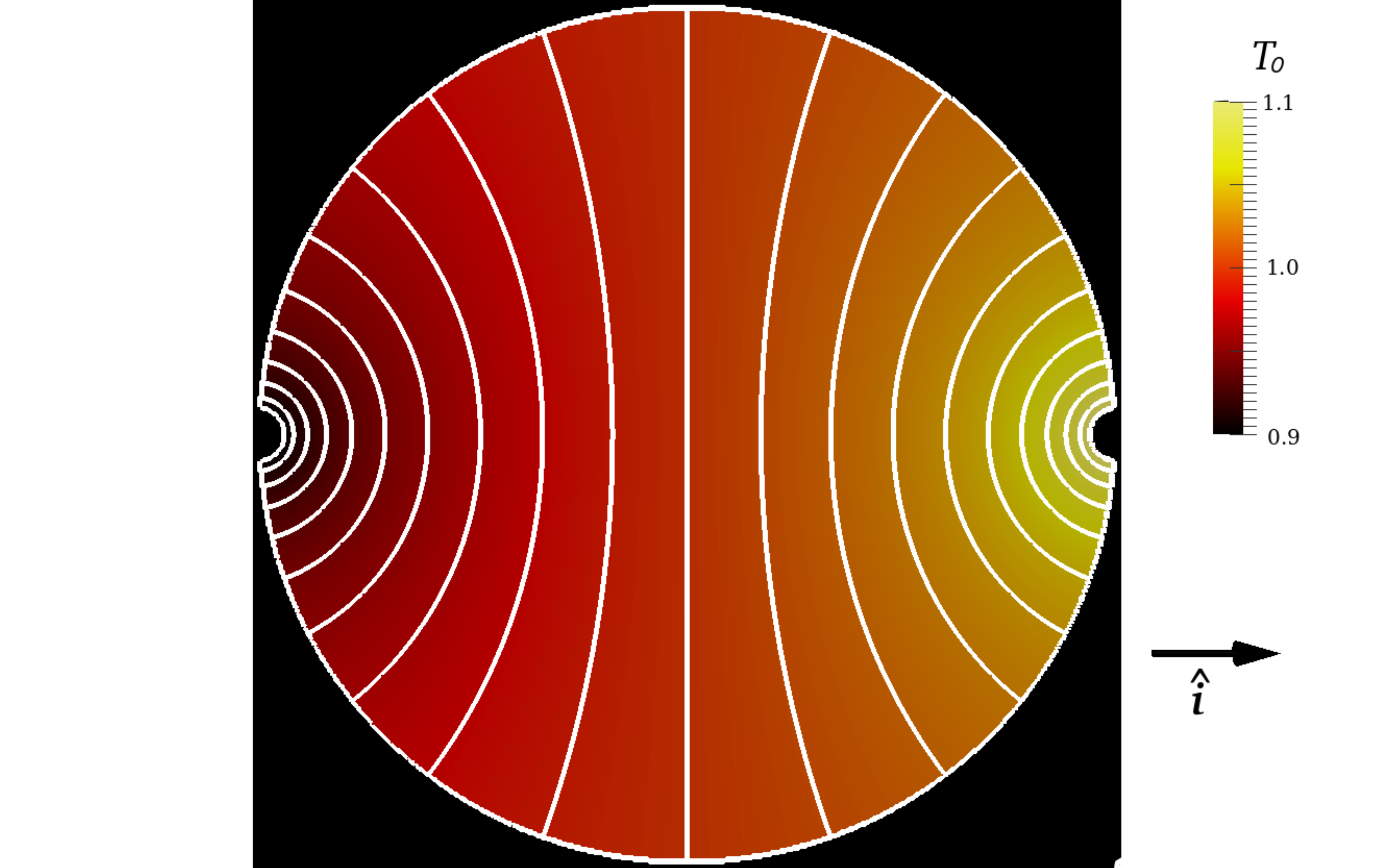}
\caption{Plot of isotherms (curves of constant $\phi$) for $T_0$, eqn.(\ref{T00}), for $\tau = 1/30$, extracting two small circles corresponding to $\Phi = 3$, close to the points $r=1$, $\theta = 0, \pi$. 
Vector $\hat{\bf e}_{x}=\hat{\bf i}$ is shown pointing in the $\theta =  0$ direction. }
\label{T_0}
\end{figure}

If the cell is in a horizontal position, so that $\alpha=0$, the equations admit a static solution $(p_0,T_0,\rho_0)$, satisfying
\begin{eqnarray}
\nabla^2 T_0  =  0, \ \ \nabla p_0  =  0, \ \ \mathcal{R} \rho_0 T_0=p_0 .
\end{eqnarray}
Laplace's equation for the circular geometry with the zero heat flux condition, can be easily solved in bipolar coordinates~\citep{happel}.
 The transformation from plane polar $(r,\theta)$ to bipolar $(\phi, \xi)$ coordinates is given by relations:
\begin{equation}
\tan{\theta} = \frac{\sin \xi}{\sinh \phi}, \ \ r = \frac{\sqrt{\sinh^2 \phi + \sin^2 \xi }}{\cosh\phi - \cos \xi}.
\end{equation}
The points at fixed temperatures are placed at $r=1$, and $\theta=0, \pi$, respectively. In bipolar coordinates, these points correspond to
 $\phi = \pm \infty$, and the unit circle corresponds to $\xi = \pi/2$, for the upper half, and $\xi= 3\pi/2$ for the lower half. Then,
 the zero heat flux condition is given by $\partial_\xi T = 0$ at $\xi = \pi/2, 3\pi/2$. 
 The solution in polar coordinates (see figure \ref{T_0}) reads,
\begin{equation}
T_0=1+ \tau  \mbox{Arctanh} \left( \frac{2r \cos \theta}{r^2 + 1} \right),
\label{T00}
\end{equation}
\noindent where $\tau =(T_h-T_c)/(2T_m \Phi)$. The parameter $\Phi$ appears due to the fact that the inverse hyperbolic tangent diverges at
 $\pm1$; the solution cannot satisfy fixed temperature conditions at the points proposed, but rather at small circles around the temperature
 sources corresponding to the values $\phi = \pm \Phi$; for example, for $\Phi=5$, the corresponding circles would be of about 1\% of the cell's
 diameter. This apparent restriction -in fact- models the finite size of the tips of the cold and hot temperature sources in the
 experiment (see figure \ref{T_0}).
\section{Perturbation analysis}
If the cell is slightly inclined (small $\alpha$), the effective gravity parameter satisfies $\rho \sin(\alpha) \ll 1$, and one can assume
 that the temperature, pressure, density and velocity fields can be expanded as power series in $\epsilon = \sin(\alpha)$. Substituting into
 eqns.(\ref{cont}-\ref{ener}) leads to an infinite set of coupled differential equations.
To zeroth order, the (static) solution corresponds to that obtained it the previous section. Notice that the velocity field starts at order
 one, ${\bf u} = \epsilon{\bf u}_1 + \epsilon^2{\bf u}_2 + \cdots$. We write only the first order problem:
\begin{eqnarray}
 \partial_t \rho_1  & = & -\nabla \cdot (\rho_0 {\bf u}_1) \label{cont_1} \\
\rho_0  \partial_t {\bf u}_1 & = & - \mathcal{M}^{-2} \nabla (\rho_0 T_1+\rho_1 T_0) + \mathcal{G} \left[ \nabla^2  {\bf u}_1  
  + \frac{\mu}{\eta} \nabla \nabla \cdot {\bf u}_1 \right] + \rho_0 \hat{\bf e}_x \label{mom_1} \nonumber\\
 \rho_0\partial_t T_1    &=& - \rho_0{\bf u}_1 \cdot\nabla T_0 + \frac{\mathcal{G}}{\mathcal{P}}  \nabla^2 T_1 - \mathcal{R}
 \rho_0 T_0 \nabla \cdot {\bf u}_1 \label{ener_1} 
\end{eqnarray}
To this order, a static solution (${\bf u}_1 = {\bf 0}$, $\partial_t T_1=0$ and $\partial_t \rho_1=0$) would require that $\nabla^2 T_1=0$
 and $ \mathcal{M}^{-2} \nabla p_1= \mathcal{R}\rho_0 \hat{\bf e}_{x}$, since $p_1 =  \mathcal{R} (\rho_0 T_1+\rho_1 T_0)$. As $\rho_0$ is a
 function of $(r,\theta)$,this equation for $p_1$ cannot be satisfied, confirming the absence of a conductive state regardless of the orientation of the inclined cell (except $\alpha=0$).
\section{Analytic approximation} 
\begin{figure}
\centering
\includegraphics[scale=0.23]{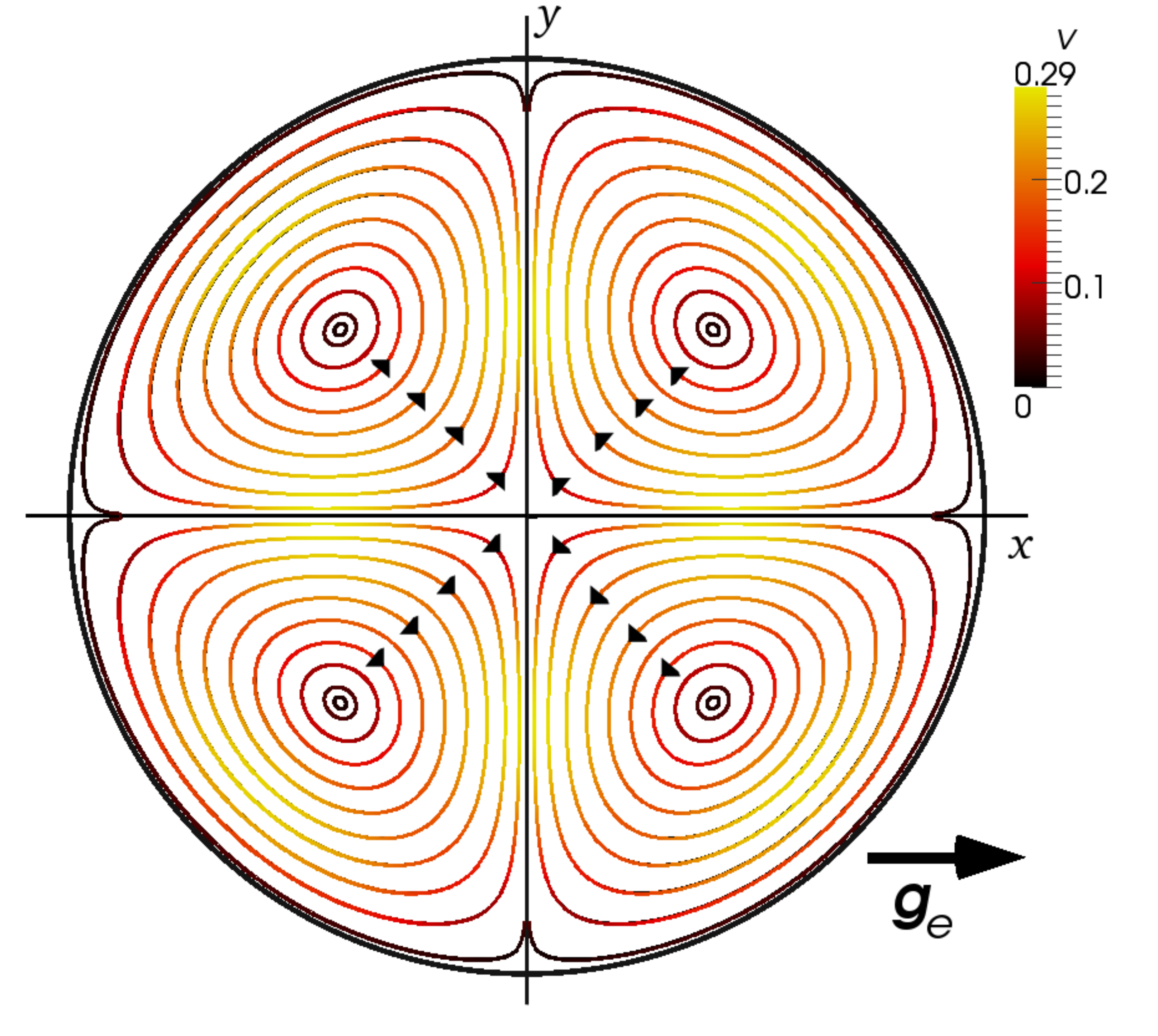}
\caption{Streamlines of the flow field showing the four rolls structure, eqns.(\ref{s1}) and (\ref{s2}),
 for $\tau p_0/(96\mathcal{R}\mathcal{G})=1$.} 
\label{analit}
\end{figure}

Consider eqn.(\ref{mom_1}) for a steady state. Taking the curl of each term, one arrives at
\begin{equation}
 \nabla^2 ( \nabla \times {\bf u}_1 ) + \nabla \times \left( \frac{\rho_0}{\mathcal{G}} \hat{\bf e}_{x} \right)= 0.
\label{mom_2}
\end{equation}
Equation (\ref{cont_1}) reduces to
\begin{equation}
 \nabla \cdot \left( \frac{\rho_0}{\mathcal{G}} {\bf u}_1  \right) = 0.
\label{cont_2}
\end{equation}
In polar coordinates ${\bf u}_1 = u \hat{\bf e}_r + v \hat{\bf e}_{\theta}$, and $\hat{\bf e}_{x}= \hat{\bf e}_r \cos{\theta} - 
\hat{\bf e}_{\theta} \sin{\theta}$. Equations (\ref{mom_2}-\ref{cont_2}) represent two inhomogeneous scalar equations for
 $u$ and $v$ that can be written as 
\begin{equation}
 (r \rho_0 u)_r + (\rho_0 v)_{\theta}= 0,
\label{C}
\end{equation}
\noindent and
\begin{eqnarray}
&& u_{\theta \theta \theta} + r^2 u_{rr \theta} - r u_{r \theta} + u_{\theta}
- r^3 v_{rrr} -2r^2 v_{rr}  + rv_r \nonumber \\
&& - r v_{r \theta\theta} - v_{\theta\theta} - v =- r^2 ( r [\frac{\rho_0}{\mathcal{G}}]_r\sin{\theta} +  [\frac{\rho_0}{\mathcal{G}}]_{\theta}\cos{\theta} ).
\label{M}
\end{eqnarray}
Here, subindices $r$ and $\theta$ represent partial derivatives. To construct an approximate solution, that remains finite in the domain and
 preserves the appropriate symmetries, one first solves the homogeneous case (without $\rho_0$).
Due to symmetry considerations, solutions are obtained for $\theta$ in the interval $[0,\pi/2]$ with $v=0$ at $\theta = 0, \pi/2$.
 The result is that a divergence free velocity field has a form in which $u$ and $v$ can be expressed as the following series
\begin{eqnarray}
u &=& \sum_{n = 0}^{\infty} u_n(r) \cos{2n\theta}, \nonumber \\
v &=& \sum_{n = 0}^{\infty} v_n(r) \sin{2n\theta}, \nonumber
\end{eqnarray}
\noindent where the coefficients are given by
\begin{eqnarray}
u_n &=& A_n r^{2n+1} + B_n r^{2n-1}, \nonumber \\
v_n &=& -\frac{n+1}{n} A_n r^{2n+1} - B_n r^{2n-1},
\nonumber
\end{eqnarray} 
\noindent such that $B_0 =0$, and $A_n$ and $B_n$ are sets of constants to be determined.\\
To complete this flow field, a particular solution of eqn.(\ref{M}) is required.
 Since $\rho_0 = p_0/(\mathcal{R} T_0)$, and $T_0$ is given by eqn.(\ref{T00}),  it can be expanded in powers of $r$ 
\begin{eqnarray}
\rho_0 &=& \frac{p_0}{\mathcal{R}} \left[ 1 - 2\tau r \cos{\theta} +
4\tau^2 r^2 \cos^2{\theta} \right. \nonumber \\ 
&+& \left. 2 \tau \left( \cos{\theta} - 4 \tau^2 \cos^3{\theta} - \frac{4 }{3}\cos^3{\theta}  \right)r^3+...\right]
\label{R}
\end{eqnarray}
Using the series solution and this expansion in eqn.(\ref{M}), a special closed solution 
that satisfies the no-slip boundary conditions can be obtained, up to order $r^5$:
\begin{eqnarray}
 u_r &=& \frac{ \tau p_0}{96\mathcal{R}\mathcal{G} } \left( 2 r^2 -1 -r^4 \right) r \cos{2 \theta} 
\label{s1}\\
 u_{\theta} &=& - \frac{ \tau p_0}{96\mathcal{R}\mathcal{G} } \left( 4 r^2 -1 - 3r^4 \right) r \sin{2 \theta} \label{s2}
\end{eqnarray}
Clearly, the velocity field vanishes on the rim and at the center of the cell; also in the fluid domain, stagnation points are located
 at $\theta = \pi/4, 3\pi/4, 5\pi/4, 7\pi/4$, and lie on the circle with radius $r = \sqrt{3}/3$. There are no adjustable parameters.\\ 
This divergence free velocity field, shown in figure \ref{analit}, is remarkably close to the numerical result in which the density varies
 within the cell; this can be understood by realizing that regions in which the density varies, the velocity is almost everywhere
 perpendicular to the density gradient.
If the cell is tilted at an angle of one degree ($\sin \alpha=0.017$), the analytical solution predicts a maximum speed within the cell 
of $0.03 cm/s$, if the source temperatures are $48^{\circ} C$ and $8^{\circ} C$, respectively.
It must be noticed that the analytical solution corresponds to a cell in almost horizontal position, hence it cannot be compared directly with the experimental observation for a cell
in the vertical position, shown in figure \ref{E}. We conjecture that the analytical solution
for an almost horizontal cell may be stable as the effect of gravity becomes negligible. 

\section{Numerical solution}
\begin{figure}
(a)
\includegraphics[scale=0.3]{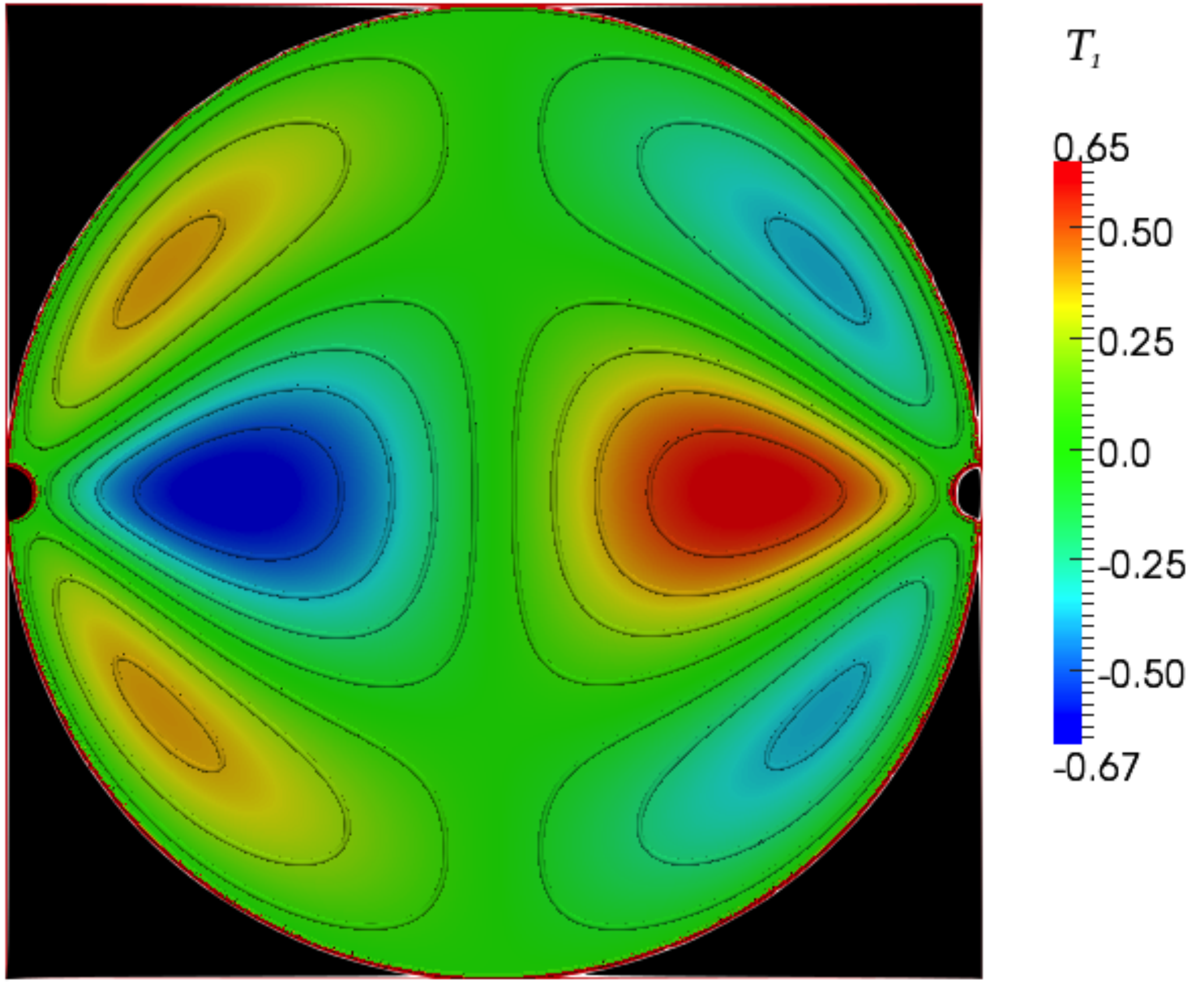}
(b)
\includegraphics[scale=0.3]{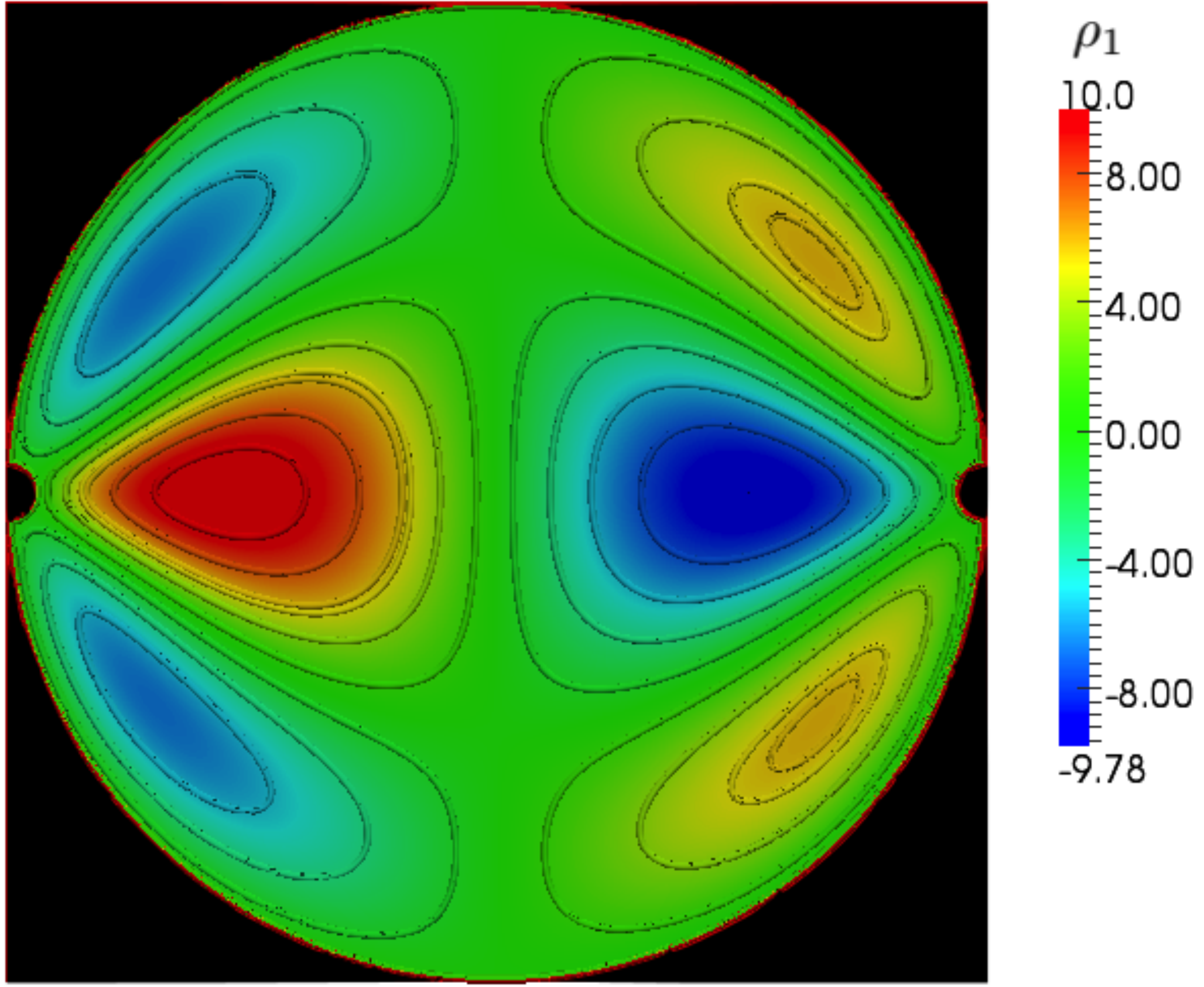}
\caption{Numerical steady state solution for $T_1$ (a) and $\rho_1$ (b).} 
\label{num}
\end{figure}

The analytical approximation to the velocity field, eqns.(\ref{s1}) and (\ref{s2}), where used as initial conditions to solve numerically the
 perturbation eqns.(\ref{mom_1}-\ref{ener_1}). The temperature and density perturbations, $T_1$ and $\rho_1$, where considered zero initially.
 No-slip and zero heat flux conditions where imposed on ${\bf u}_1$ and $T_1$ respectively. At the heat source and sink, $T_1$ was fixed to
 zero. Time iterations where performed until the perturbed fields reached an approximate steady state. Figure \ref{num} shows the steady state
 numerical solutions for $\rho_1$ and $T_1$ corresponding to parameters $\mathcal{M}= 0.01$, $\mathcal{P}=7$, $\mathcal{R}=1$; and conditions
 $T_c=0.9$, $T_h=1.1$, and $\Phi=3.5$.
Solutions show the expected behaviour. As fluid is convected away from the heat source towards the center of the cell it heats up the region
 in front of the source. Fluid convected away from the heat sink towards the center cools down the region in front of the sink. This is shown
 by the local maximum and minimum of $T_1$ located on the line connecting the source and sink; not an intuitive expectation. As fluid is
 drawn from the meridional region back to the heat source it should decrease the temperature on the regions close to the boundaries and the
 source. Fluid drawn back to the heat sink, should do the opposite. This is also shown in figure \ref{num} by the local maxima and minima
 of $T_1$ close to the boundaries of the cell. Correspondingly, density perturbations $\rho_1$ show the opposite behavior.
We emphasize that the analytical initial condition for the velocity ${\bf u}_1$ represents an incompressible flow. Therefore it can not
 satisfy eqn.(\ref{cont_1}) at steady state. As contour lines of $\rho_0$ (which correspond to contour lines of $T_0$, shown in
 figure \ref{T_0} are not everywhere perpendicular to the stream lines of  ${\bf u}_1$ (see figure \ref{analit}), the term
 ${\bf u}_1 \cdot \nabla \rho_0$ can not be zero everywhere. Nevertheless, for a small temperature difference between source
 and sink, $\rho_0$ is almost flat everywhere except in the vicinity of the source and sink, which are regions where small fluid
 compressions and expansions occur. Therefore, for the analytic initial velocity ${\bf u}_1 \cdot \nabla \rho_0 \simeq 0$ almost
 everywhere, and the numerical procedure only slightly modifies it close to the source and sink, which is rather interesting and
 worth pointing out, as incompressibility, while not been imposed, seems to be a very precise approximation, not all that foreign
 to the Boussinesq assumptions.
\section{Conclusions}
This paper addresses the general problem of thermal convection due to temperature differences in a fluid system, in which the body force
 -gravity- can be varied. Relevant new findings of the present analysis are: a) The chosen (circular and planar) geometry is specially
 suited to show that fluid motion is always present, if there is a temperature gradient and a non zero gravity (external force); regardless of the value of all dimensionless parameters, including Rayleigh's number
 \citep{batchelor,landau,gershuni}. b) A static state with pure heat conduction is possible if {\sl gravity} is strictly zero, and it is
 probably unstable, as -implicitly- the present approach using perturbation theory suggests. c) Avoiding the Boussinesq approximation,
 as arbitrarily small {\sl gravity} makes it unsuitable, leads to a set of equations that allow a closed analytical solution of the
 velocity field, showing four symmetrical rolls, one on each of the quadrants of the system, the signature of convection cells. d)
 Numerical solutions of the perturbed equations show steady state distributions of the perturbed temperature and density consistent
 with the analytical velocity field. Experiments, and a numerical solution of the full hydrodynamic problem are been completed;
 all preliminary results are consistent with the present analysis, and will be published elsewhere.
 
\section{Acknowledgment} 
RPF enjoyed a Sabbatical Fellowship from the Direcci\'on General de Asuntos del Personal Acad\'emico-UNAM, and the hospitality at the PMMH-\'Ecole Sup\'erieure de Physique et Chimie Industrielles (ESPCI), Paris, France. Thanks also for interesting comments from L. S. Tuckerman and J. E. Wesfried.


\end{document}